 \documentclass[]{revtex4-2}

\usepackage{pstricks,pst-all}

\usepackage{textcomp}
\usepackage{gensymb}
\usepackage{subcaption}
\usepackage{wrapfig}
\usepackage{graphicx}
\usepackage{mathtools}
\usepackage{tabularx}
\usepackage{amssymb}
\usepackage{amsmath, nccmath}
\usepackage{commath}
\usepackage{amsthm}
\usepackage{bm}
\usepackage{lineno}
\makeatletter
\newcommand*{\rom}[1]{\expandafter\@slowromancap\romannumeral #1@}
\makeatother
\usepackage{lipsum}
\makeatletter
\def\ps@pprintTitle{%
 \let\@oddhead\@empty
 \let\@evenhead\@empty
 \def\@oddfoot{}%
 \let\@evenfoot\@oddfoot}
\makeatother

\begin{document}

\title{Randomness in quantum random number generator from vacuum fluctuations with source-device-independence}

\author{Megha Shrivastava}
\author{Mohit Mittal}
\author{Isha Kumari}
\author{Venkat Abhignan}
 \email{yvabhignan@gmail.com}
\affiliation{Qdit Labs Pvt. Ltd., Bengaluru - 560092, India}

\begin{abstract}
The application for random numbers is ubiquitous. We experimentally build a well-studied quantum random number generator from homodyne measurements on the quadrature of the vacuum fluctuations. Semi-device-independence in this random number generator is usually obtained using phase modulators to shift the phase of the laser and obtain random sampling from both X and P quadrature measurements of the vacuum state in previous implementations. We characterize the experimental parameters for optimal performance of this source-device independent quantum random number generator by measuring the two quadratures concurrently using two homodyne detectors. We also study the influence of these parameters on randomness, which can be extracted based on Shannon entropy and von Neumann entropy, which correspond to an eavesdropper listening to classical and quantum side information, respectively.
\end{abstract}

\maketitle

\section{INTRODUCTION}

Random numbers are employed in many different contexts, such as simulations, cryptography and fundamental science \cite{RevModPhys.89.015004}. Pseudo-random number generators are based on deterministic methods \cite{10.5555/270146} are usually used to effectively and efficiently provide random numbers \cite{hörmann2004automatic}. However, because their output solely depends on a particular algorithm and the original seed, it can be proven to have an inherent periodicity, making it predictable with sufficient computing power. The property of intrinsic randomness in the random numbers is critical for most applications. Security will suffer if cryptographic keys generated from pseudo-random numbers exhibit predictability. 

Quantum random number generators (QRNG) are one of the most developed quantum optics-based technologies exploiting the intrinsic randomness of quantum phenomenon \cite{stipvcevic2011quantum,Ma2016}. Due to the challenges in measuring the quantum phenomenon, most QRNG implementations have been restricted to a relatively low rate. For example, the maximum counting rate of single-photon detectors, which is typically below 100 MHz, limits the speed of single-photon-detection-based QRNG \cite{SINGLE1,SINGLE2,SINGLE3,SINGLE4,SINGLE5}. A continuous-variable QRNG scheme taking advantage of homodyne measurements \cite{collett} of quadrature fluctuations in the vacuum field efficiently obtains a higher random number sampling rate \cite{Yuen:83,Trifonov}. It utilizes the coherent detection technique, which eliminates the restriction of detector dead time by substituting high-performance homodyne photodetectors for single-photon detectors, which is primarily responsible for significantly improving randomness generation performance \cite{Gabriel2010,10.1063/1.3597793,PhysRevApplied.3.054004}. Field-programmable-gate-array (FPGA) implementations of information-theoretically secure Toeplitz randomness extractor have been shown to extract random numbers in real-time at GB/s speed using this method \cite{Zhang2016,10.1063/1.5078547}. 

Further, QRNG implementations that can produce randomness verified with source-independence (SI) are considered more secure \cite{PhysRevLett.118.060503,PhysRevApplied.12.034017}. Recently, it was also shown that SI-QRNG can generate random numbers up to GB/s speeds \cite{Avesani2018,Xu_2019,PhysRevA.99.062326,Cheng:22}. SI-QRNG uses homodyne or heterodyne detection to measure randomly two quadrature observables X and P of an input untrusted quantum state (vacuum state), where the phase of the continuous-wave laser selects the quadrature. This ensures the security of the generated random numbers, even with an untrusted source. However, to alter the phase output of the continuous-wave laser, the homodyne-based and heterodyne-based SI-QRNG protocols require the addition of a phase modulator. In particular, the homodyne-based CV-SI-QRNG protocols require external initial randomness, making the SI-QRNG setup more complex. 

Recently, by taking advantage of a fully reliable beam splitter, concurrently, phase differences of $0$ and $\pi/2$ were applied between the vacuum source and laser signal to determine the X and P quadrature measurements separately by using two homodyne detectors \cite{Cheng:22}. We implemented a similar setup to measure the two quadratures simultaneously, and we characterize the experimental conditions for optimal performance of this quantum random number generator in Sec. 2. Further, we examine how these characteristics affect randomness, which is obtained using von Neumann and Shannon entropies in Sec. 3.

\section{EXPERIMENTAL SETUP}
Our experimental setup comprises a fully fibre-connected structure with commercially available components for randomness generation. As shown in Fig. 1, the laser signal (TeraXion PS-NLL-1550) is initially split using a $1\times2$ balanced coupler. The laser signal in each arm is phase shifted using PM1 and PM2 by imparting phases 0 and $\pi/2$ to measure X and P quadratures concurrently. Further,  in each arm, two output beams are produced with balanced power by the interference of the vacuum state and phase-modulated laser signal on a symmetric beamsplitter (50:50 BS), which are then fed to two balanced homodyne photodetectors (BHD1 and BHD2, Thorlabs balanced photodetector 1.6 GHz module PDB480C-AC). Then, it is fed to an analog-to-digital converter (ADC) and high-speed field-programmable gate array (P0435 Cyclone V SE SoC ADC-SoC 5CSEMA4 Cyclone$^\text{®}$ V SE FPGA + MCU/MPU SoC Evaluation Board) for the extraction of random numbers in post-processing. 

\begin{figure}
    \centering
    \includegraphics[width=17cm]{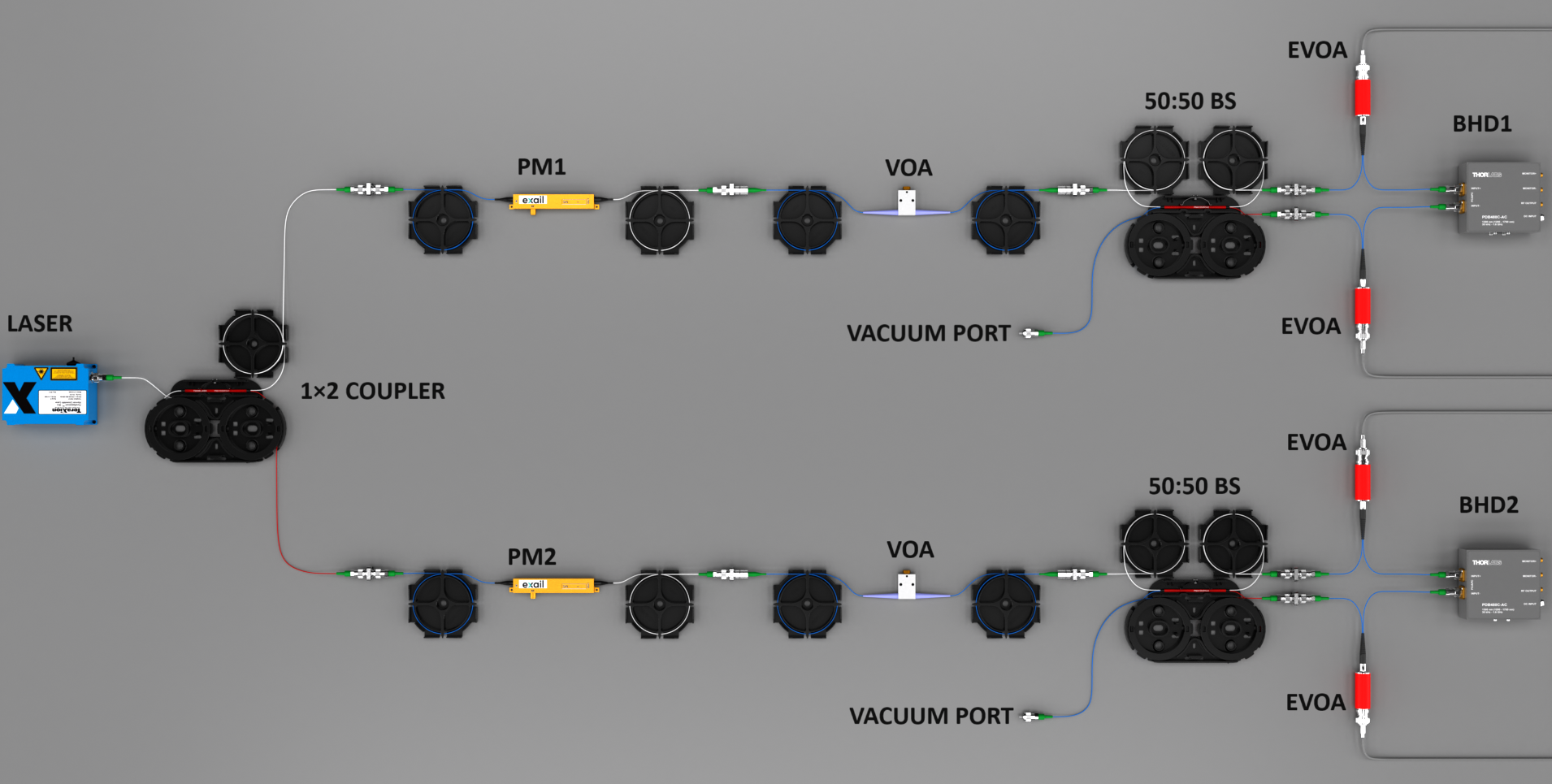}
    \caption{Design of the quantum random number generator. PM is phase modulator, VOA is variable optical attenuator, BS is beam splitter, EVOA is electronic VOA and BHD is balanced homodyne detector.}
    \label{fig:enter-label}
\end{figure}

It is impossible to eradicate classical noise $E$ completely, and it will also be incorporated into the recorded raw data consisting of quantum noise $Q$. For optimal performance of the random number generator, the quantum noise needs to exceed the classical noise by 10 dB to yield an ideal number of random bits per sample \cite{8444971}, defined as QCNR (Quantum to classical noise ratio). The voltage $V$ measured by the homodyne detector has a Gaussian distribution $P(V)$ such as \cite{10.1063/1.5078547} \begin{equation}
    P(V)=\frac{1}{\sqrt{2\pi(\sigma^2_{Q}+\sigma^2_{E}+2(\delta/12)^2)}}\exp{\left(-\frac{V^2}{2(\sigma^2_{Q}+\sigma^2_{E}+2(\delta/12)^2))}\right)},
\end{equation} consisting of vacuum signal (with variance $\sigma^2_{Q}$ for $Q=x,p$ at two BHDs) and classical noise  (with variance $\sigma^2_{E}$). When the laser is turned off, the variance $\sigma^2_{E}$ of the sampled raw data has a non-zero value, typically attributed to the classical noise caused by the electromagnetic disturbance, the temperature fluctuations, and the inherent flaws in the experimental setup. The ADC used has an impact of quantization error $\delta=2R/(2^n)$ depending on the range of input voltage $R$ and sampling precision $n$. 

When the laser is turned on, the quantum signal with vacuum fluctuations starts dominating, and the optimal performance for maximum $\sigma^2_{Q}$ can be obtained by varying the power of the laser \cite{Xu_2019}. The voltage $V$ is recorded by varying the power of the laser in Fig. 2. With the phase 0($\pi/2$) to PM1(PM2), the variance of voltage ($V^2$) measured at BHDs increases linearly with an increase in power of the laser. In the range of 0 mW to 9 mW, the saturation reached at a laser power of 9 mW for BHD1 associated with PM1 and BHD2 associated with PM2 in Fig. 2(a) and (b), respectively. The quantum to classical noise relation (QCNR=$10\log_{10}{V_Q/V_E}$) can be studied as observed in Fig. 3, which again saturates at 9 mW, giving the optimal value of around 10 dB as observed.

 \begin{figure}[ht]
\centering
\begin{subfigure}{0.46\textwidth}
\includegraphics[width=1\linewidth, height=5cm]{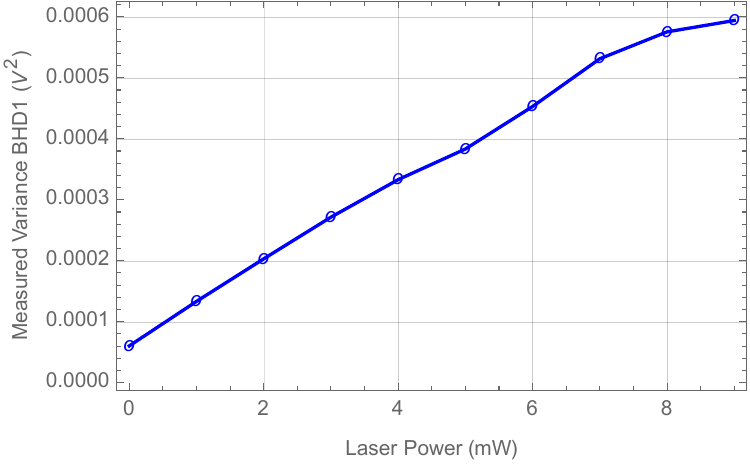} 
\caption{Laser power (mW) vs Variance of voltage ($V^2$) for BHD1.}

\end{subfigure}
\begin{subfigure}{0.46\textwidth}
\includegraphics[width=1\linewidth, height=5cm]{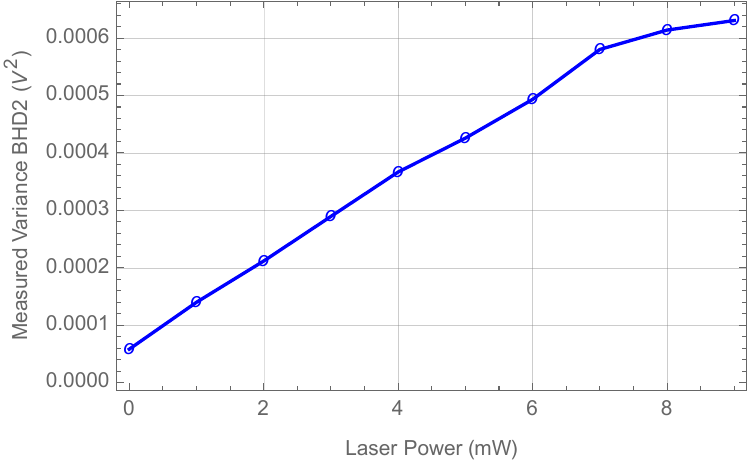}
\caption{Laser power (mW) vs Variance of voltage ($V^2$) for BHD2.}

\end{subfigure}
 
\caption{ As a function of laser power, these figures display the voltage variance of the sampled raw data for the BHDs. The behaviour of the voltage variance is relatively linear in the range of 0 to 7 mW and saturates at 9 mW where we obtain the optimal performance.}

\end{figure}

 \begin{figure}[ht]
\centering
\begin{subfigure}{0.46\textwidth}
\includegraphics[width=1\linewidth, height=5cm]{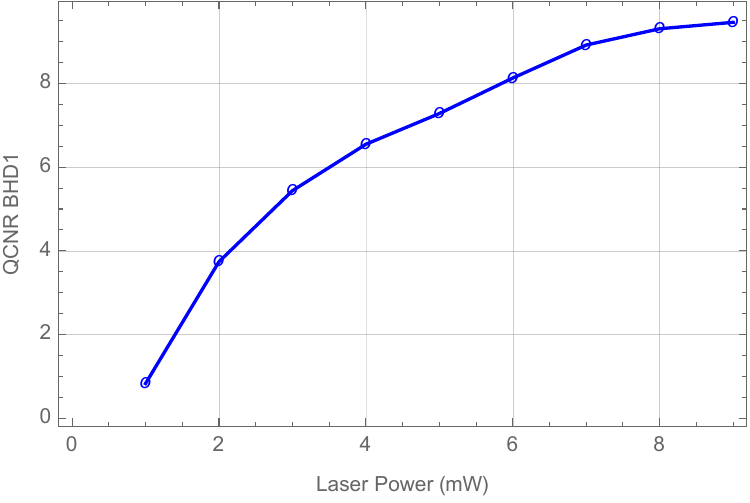} 
\caption{QCNR for the variance $V_x$ at BHD1.}

\end{subfigure}
\begin{subfigure}{0.46\textwidth}
\includegraphics[width=1\linewidth, height=5cm]{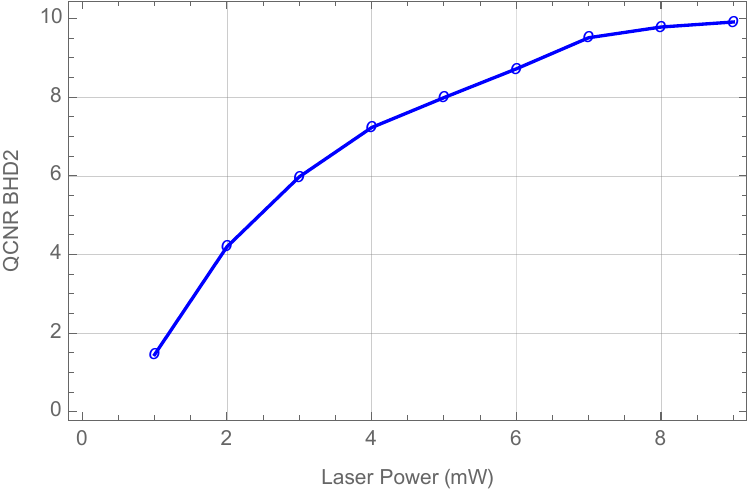}
\caption{QCNR for the variance $V_p$ at BHD2.}

\end{subfigure}
 
\caption{QCNR vs Laser power. The QCNR of sampled raw data at the BHDs shows saturation at 9 mW.}

\end{figure}

The approach to studying the optimal parameters is from an RF spectrum analyzer. RF spectrum can be used to perform the frequency domain measurements to measure the power spectral density of the vacuum fluctuations relative to the classical noise. These readings show that the measured power spectral density of the vacuum fluctuations is maximum when the laser has the power of 9 mW, as shown in Fig. 4. The power spectral density increases with an increase in laser power, and the power spectral density measured by the balanced homodyne receivers reaches saturation at a laser power of 9 mW, resulting in the optimal values. The power spectral density is recorded for the vacuum fluctuations relative to the classical noise at the optimal 9 mW power of the laser in both the BHDs, as illustrated in Fig. 5, clearly differentiating them.  
\begin{figure}[ht]
\centering
\begin{subfigure}{0.46\textwidth}
\includegraphics[width=1\linewidth, height=5cm]{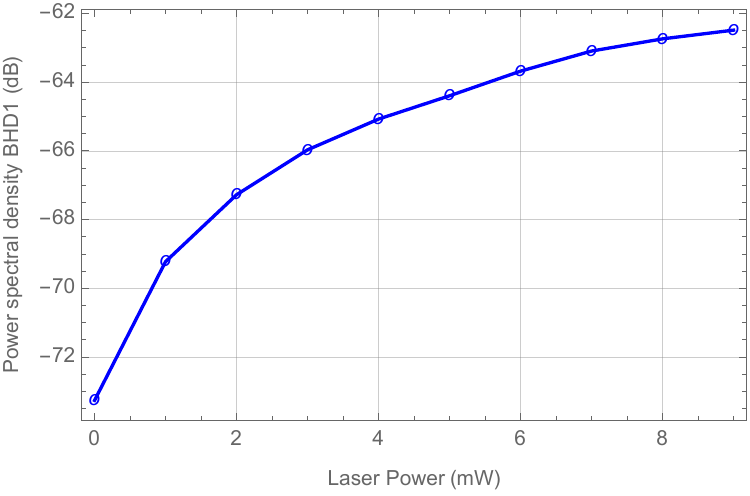} 
\caption{The power spectral density at BHD1.}

\end{subfigure}
\begin{subfigure}{0.46\textwidth}
\includegraphics[width=1\linewidth, height=5cm]{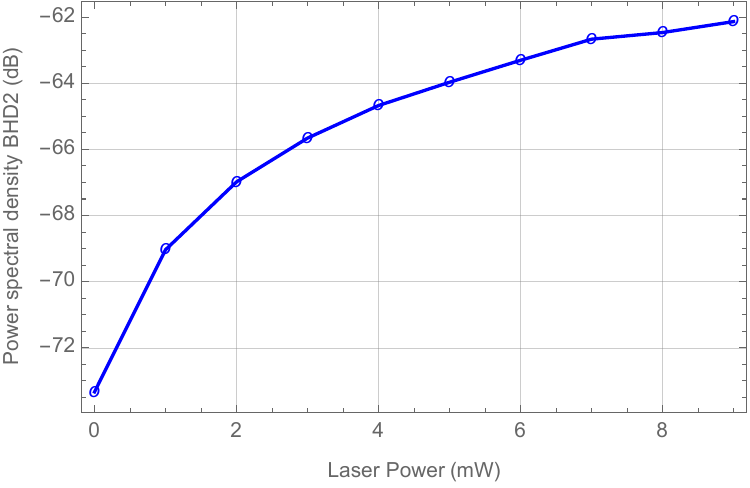}
\caption{The power spectral density at BHD2.}

\end{subfigure}
 
\caption{The power spectral density vs Laser power. It shows that power spectral density in both the BHDs saturate at a laser power of 9 mW.}

\end{figure}
\begin{figure}[ht]
\centering
\begin{subfigure}{0.46\textwidth}
\includegraphics[width=1.1\linewidth, height=7cm]{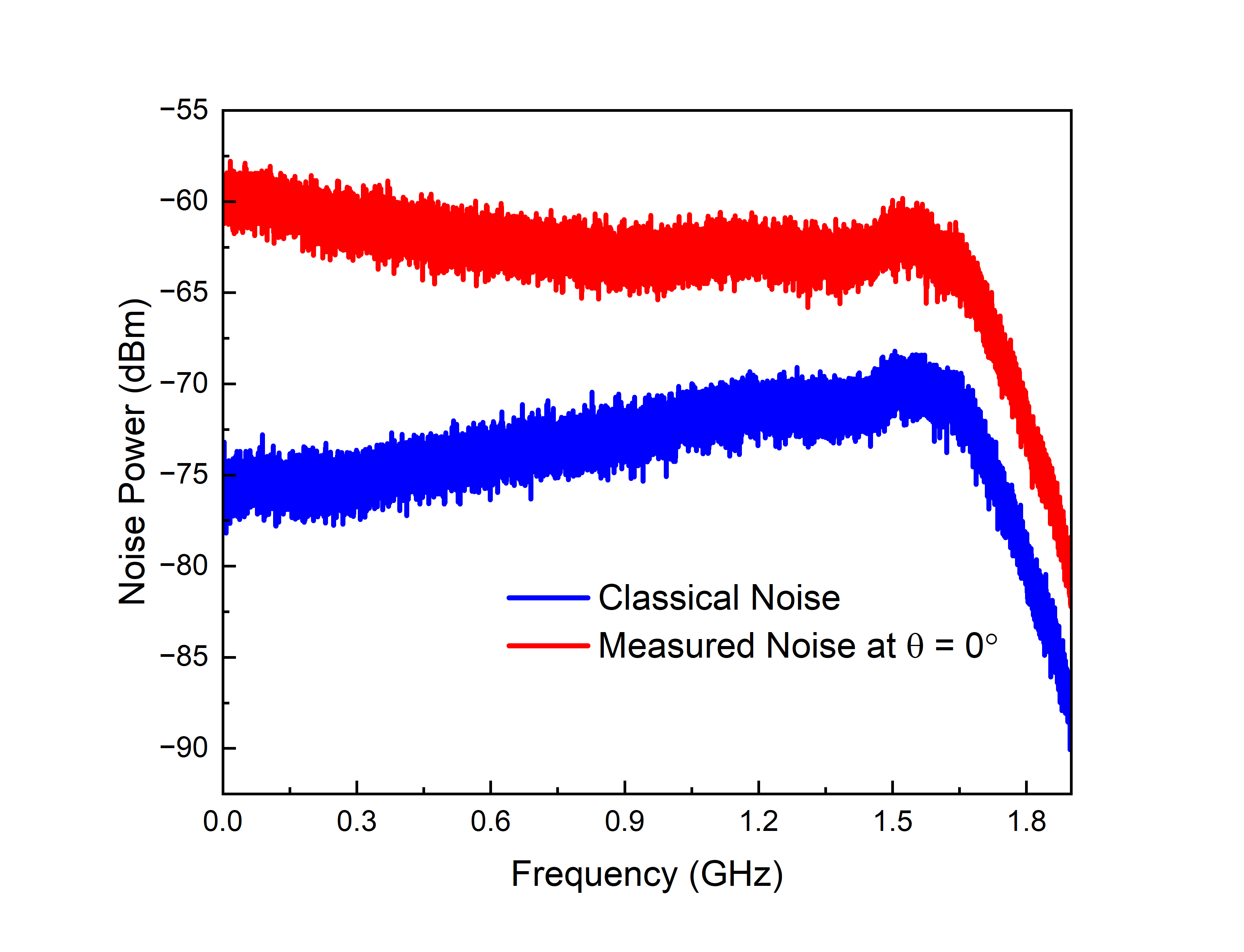} 
\caption{The power spectral density at BHD1.}

\end{subfigure}
\begin{subfigure}{0.46\textwidth}
\includegraphics[width=1.1\linewidth, height=7cm]{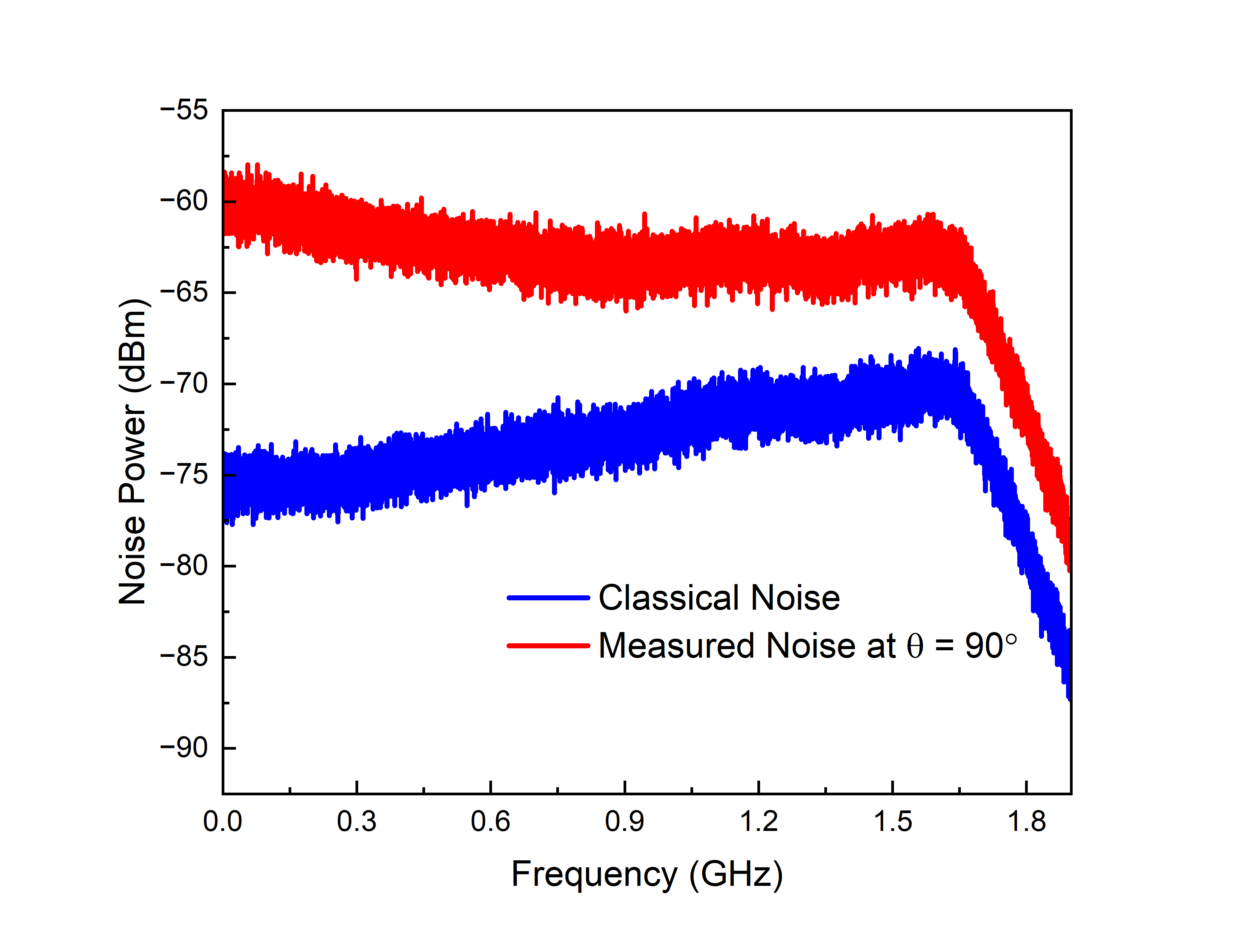}
\caption{The power spectral density at BHD2.}

\end{subfigure}
 
\caption{The power spectral density for the vacuum fluctuations in relation to the classical noise at 9 mW laser power.}

\end{figure}

\section{Extractable randomness with source-independence} 

Based on Ref. \cite{Xu_2019}, we measure the extractable randomness based on the theory of extremality for Gaussian states of vacuum \cite{PhysRevLett.96.080502,PhysRevLett.97.190503}. An estimate for the bound of extractable randomness is given by the covariance matrix ($CM$) of these two measured quadratures $X$ and $P$ of the quantum state, which can be written as 

\begin{equation}
	\label{gammaA}
	\begin{split}
		CM = \left( {\begin{array}{*{20}{c}}{{\sigma^2_{x}}}&{c}\\
				{c}&{{\sigma^2_{p}}}\end{array}} \right).
	\end{split}
\end{equation}
% \fi
Here ${{\sigma^2_{x}}}$ and ${{\sigma^2_{p}}}$ the variances of $X$ and $P$ quadratures at the two BHDs, and ${c}$ is the co-variance between $X$ and $P$ quadratures. Notably, the values of ${{\sigma^2_{x}}}$ and ${{\sigma^2_{p}}}$ are nearly equal as can be seen in Fig. 2. As in the case of the security analysis in the homodyne-based SI-QRNG \cite {Xu_2019}, the lower bound of the extractable randomness per measurement $R$ conditioned on the presence of an eavesdropper can be obtained as

\begin{equation}
	\label{Rdis1}
	\begin{split}
		R \ge H_{\text{min}} - S,
	\end{split}
\end{equation} where $H_{\text{min}}$ is the Shannon entropy of quadrature $X$ from which random numbers are generated and von Neumann entropy $S$ corresponds to quantum side information dependent on variance of quadratures $X$ and $P$. Previously, this conditional min-entropy $H_{\text{min}}$ was used to determine the random bits that can be generated from each sample, assuming that an eavesdropper with full knowledge of setup can listen to only the classical noise \cite{PhysRevApplied.3.054004}. And the $H_{\text{min}}$ was calculated as \cite{10.1063/1.5078547} \begin{equation}
    H_{\text{min}}=-\log_2\left[\text{erf}\left(\frac{\delta}{2\sqrt{2\sigma^2_{x}}}\right)\right].
\end{equation} In the source-independent scenario the  von Neumann entropy $S$ has a Holevo's bound \cite{Xu_2019} that can be computed as \begin{equation}
    S = [(\lambda  + 1)/2]{\log _2}[(\lambda  + 1)/2] - [(\lambda  - 1)/2]{\log _2}[(\lambda  - 1)/2], \end{equation} where $\lambda = \sqrt {\det \left({CM} \right)}  = \sqrt {{\sigma^2_{x}}{\sigma^2_{p}} - {c^2}}$ from which ($H_{\text{min}}$-S/.14)\% of the sample (for 14bit ADC) can be extracted to give $H_{\text{min}}-S$ random bits. 

Considering the optimal situation as discussed in the earlier section, at the laser power of 9 mW, we measure the variance from sampled raw data as $5.9394 \times 10^{-4}V^2$ with variance from classical noise as $0.6037 \times 10^{-4}V^2$ at BHD1. Similarly, at BHD2, the variance of sampled raw data is $6.3054 \times 10^{-4} V^2$, and classical noise variance is $0.5841 \times 10^{-4} V^2$. We obtain the quantization error of ADC $\delta=1.2207\times 10^{-4}$ with voltage range $R=1V$ and sampling precision $n=14$. With this, we obtain the variance of X quadrature at BHD1 as $V_x=5.3357\times 10^{-4} V^2$ and variance of P quadrature at BHD2 as $V_p=5.7213\times 10^{-4} V^2$. Using these, we obtain Shannon entropy $H_{\text{min}}=8.8897$, von Neumann entropy $S=0.3009$ considering the upper bound with $c=0$ and $61\%$ of the raw sample can be extracted to obtain random numbers. These results are close to the implementation in Ref. \cite{Xu_2019} where they experimentally obtain Shannon entropy as $8.7117$ and von Neumann entropy as $0.3366$ with 12-bit ADC.

Further, the randomness was characterized by 1GB of recorded data utilizing the 15 statistical tests offered by the National Institute of Standards and Technology (NIST SP 800-22) \cite{lawrence}. The probability $\alpha$ is a confidence threshold that is determined before the tests. $\alpha$ is the likelihood that the tests will show the obtained random number sequence is not random when, in fact, the sequence is random. In cryptography, $\alpha$ is typically valued at 0.01. Also, for these tests, a P-value represents the likelihood that a perfect random number generator would have generated a less random sequence than the sequence under test. When a test's P-value is found to be 1, it suggests that the sequence has complete randomness. When the sequence looks to be non-random, the P-value is 0. P-value $\geq\alpha$ indicates acceptance of the null hypothesis, meaning the sequence is random. The null hypothesis is rejected if the P-value is less than $\alpha$, indicating that the sequence is not random. We show that the P-values (logarithmic scale) for the 15 statistical tests are greater than $\alpha=0.01$ in Fig. 6.

\begin{figure}
    \centering
    \includegraphics[width=11cm]{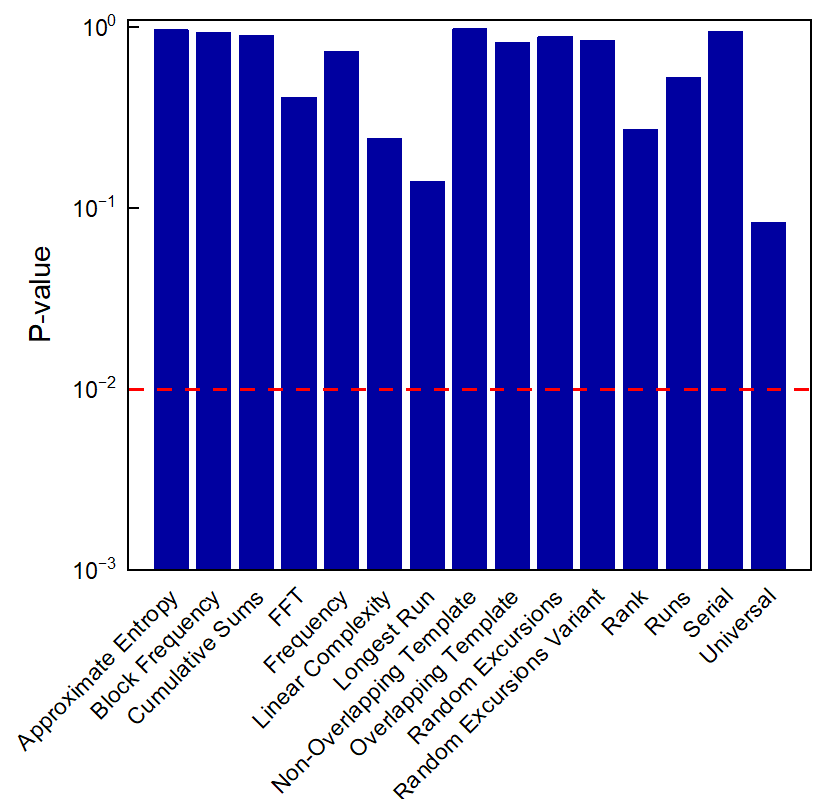}
    \caption{The NIST (National Institute of Standard and Technology) statistical test suite recordings are presented here. To pass the NIST SP800-22, the P-values that are obtained in each test need to be more than 0.01.}
    \label{fig:enter-label}
\end{figure}

\section{Conclusion}
We have implemented and optimized a quantum random number generator based on homodyne measurements of vacuum fluctuations. Similar to Ref. \cite{Cheng:22}, our approach utilizes two balanced homodyne detectors, two-phase modulators and a trusted beam-splitter to achieve semi-device independence with a concurrent sampling of X and P quadrature measurements. It is a convenient method by avoiding switching the measurements \cite{PhysRevLett.118.060503} and simultaneously measuring the two quadratures, one for raw data and the other for checking the data. We have demonstrated optimal performance in generating randomness through parameter characterization, evaluated through Shannon entropy \cite{10.1063/1.5078547} and von Neumann entropy \cite{Xu_2019}. While previously entropic uncertainty principle \cite{PhysRevLett.118.060503} was implemented to compute the lower bound for randomness \cite{Cheng:22}, we calculated using the covariance matrix of the quadratures. 
\bibliography{apssamp}% Produces the bibliography via BibTeX.

\begin{thebibliography}{28}%
\makeatletter
\providecommand \@ifxundefined [1]{%
 \@ifx{#1\undefined}
}%
\providecommand \@ifnum [1]{%
 \ifnum #1\expandafter \@firstoftwo
 \else \expandafter \@secondoftwo
 \fi
}%
\providecommand \@ifx [1]{%
 \ifx #1\expandafter \@firstoftwo
 \else \expandafter \@secondoftwo
 \fi
}%
\providecommand \natexlab [1]{#1}%
\providecommand \enquote  [1]{``#1''}%
\providecommand \bibnamefont  [1]{#1}%
\providecommand \bibfnamefont [1]{#1}%
\providecommand \citenamefont [1]{#1}%
\providecommand \href@noop [0]{\@secondoftwo}%
\providecommand \href [0]{\begingroup \@sanitize@url \@href}%
\providecommand \@href[1]{\@@startlink{#1}\@@href}%
\providecommand \@@href[1]{\endgroup#1\@@endlink}%
\providecommand \@sanitize@url [0]{\catcode `\\12\catcode `\$12\catcode `\&12\catcode `\#12\catcode `\^12\catcode `\_12\catcode `\%12\relax}%
\providecommand \@@startlink[1]{}%
\providecommand \@@endlink[0]{}%
\providecommand \url  [0]{\begingroup\@sanitize@url \@url }%
\providecommand \@url [1]{\endgroup\@href {#1}{\urlprefix }}%
\providecommand \urlprefix  [0]{URL }%
\providecommand \Eprint [0]{\href }%
\providecommand \doibase [0]{https://doi.org/}%
\providecommand \selectlanguage [0]{\@gobble}%
\providecommand \bibinfo  [0]{\@secondoftwo}%
\providecommand \bibfield  [0]{\@secondoftwo}%
\providecommand \translation [1]{[#1]}%
\providecommand \BibitemOpen [0]{}%
\providecommand \bibitemStop [0]{}%
\providecommand \bibitemNoStop [0]{.\EOS\space}%
\providecommand \EOS [0]{\spacefactor3000\relax}%
\providecommand \BibitemShut  [1]{\csname bibitem#1\endcsname}%
\let\auto@bib@innerbib\@empty
%</preamble>
\bibitem [{\citenamefont {Herrero-Collantes}\ and\ \citenamefont {Garcia-Escartin}(2017)}]{RevModPhys.89.015004}%
  \BibitemOpen
  \bibfield  {author} {\bibinfo {author} {\bibfnamefont {M.}~\bibnamefont {Herrero-Collantes}}\ and\ \bibinfo {author} {\bibfnamefont {J.~C.}\ \bibnamefont {Garcia-Escartin}},\ }\bibfield  {title} {\bibinfo {title} {Quantum random number generators},\ }\href {https://doi.org/10.1103/RevModPhys.89.015004} {\bibfield  {journal} {\bibinfo  {journal} {Rev. Mod. Phys.}\ }\textbf {\bibinfo {volume} {89}},\ \bibinfo {pages} {015004} (\bibinfo {year} {2017})}\BibitemShut {NoStop}%
\bibitem [{\citenamefont {Knuth}(1997)}]{10.5555/270146}%
  \BibitemOpen
  \bibfield  {author} {\bibinfo {author} {\bibfnamefont {D.~E.}\ \bibnamefont {Knuth}},\ }\href@noop {} {\emph {\bibinfo {title} {The art of computer programming, volume 2 (3rd ed.): seminumerical algorithms}}}\ (\bibinfo  {publisher} {Addison-Wesley Longman Publishing Co., Inc.},\ \bibinfo {address} {USA},\ \bibinfo {year} {1997})\BibitemShut {NoStop}%
\bibitem [{\citenamefont {H{\"o}rmann}\ \emph {et~al.}(2004)\citenamefont {H{\"o}rmann}, \citenamefont {Leydold},\ and\ \citenamefont {Derflinger}}]{hörmann2004automatic}%
  \BibitemOpen
  \bibfield  {author} {\bibinfo {author} {\bibfnamefont {W.}~\bibnamefont {H{\"o}rmann}}, \bibinfo {author} {\bibfnamefont {J.}~\bibnamefont {Leydold}},\ and\ \bibinfo {author} {\bibfnamefont {G.}~\bibnamefont {Derflinger}},\ }\href {https://books.google.co.in/books?id=vjOJySM27PIC} {\emph {\bibinfo {title} {Automatic Nonuniform Random Variate Generation}}},\ Statistics and Computing\ (\bibinfo  {publisher} {Springer},\ \bibinfo {year} {2004})\BibitemShut {NoStop}%
\bibitem [{\citenamefont {Stip{\v{c}}evi{\'c}}(2011)}]{stipvcevic2011quantum}%
  \BibitemOpen
  \bibfield  {author} {\bibinfo {author} {\bibfnamefont {M.}~\bibnamefont {Stip{\v{c}}evi{\'c}}},\ }\bibfield  {title} {\bibinfo {title} {Quantum random number generators and their use in cryptography},\ }in\ \href@noop {} {\emph {\bibinfo {booktitle} {2011 Proceedings of the 34th International Convention MIPRO}}}\ (\bibinfo {organization} {IEEE},\ \bibinfo {year} {2011})\ pp.\ \bibinfo {pages} {1474--1479}\BibitemShut {NoStop}%
\bibitem [{\citenamefont {Ma}\ \emph {et~al.}(2016)\citenamefont {Ma}, \citenamefont {Yuan}, \citenamefont {Cao}, \citenamefont {Qi},\ and\ \citenamefont {Zhang}}]{Ma2016}%
  \BibitemOpen
  \bibfield  {author} {\bibinfo {author} {\bibfnamefont {X.}~\bibnamefont {Ma}}, \bibinfo {author} {\bibfnamefont {X.}~\bibnamefont {Yuan}}, \bibinfo {author} {\bibfnamefont {Z.}~\bibnamefont {Cao}}, \bibinfo {author} {\bibfnamefont {B.}~\bibnamefont {Qi}},\ and\ \bibinfo {author} {\bibfnamefont {Z.}~\bibnamefont {Zhang}},\ }\bibfield  {title} {\bibinfo {title} {Quantum random number generation},\ }\href {https://doi.org/10.1038/npjqi.2016.21} {\bibfield  {journal} {\bibinfo  {journal} {npj Quantum Information}\ }\textbf {\bibinfo {volume} {2}},\ \bibinfo {pages} {16021} (\bibinfo {year} {2016})}\BibitemShut {NoStop}%
\bibitem [{\citenamefont {Jennewein}\ \emph {et~al.}(2000)\citenamefont {Jennewein}, \citenamefont {Achleitner}, \citenamefont {Weihs}, \citenamefont {Weinfurter},\ and\ \citenamefont {Zeilinger}}]{SINGLE1}%
  \BibitemOpen
  \bibfield  {author} {\bibinfo {author} {\bibfnamefont {T.}~\bibnamefont {Jennewein}}, \bibinfo {author} {\bibfnamefont {U.}~\bibnamefont {Achleitner}}, \bibinfo {author} {\bibfnamefont {G.}~\bibnamefont {Weihs}}, \bibinfo {author} {\bibfnamefont {H.}~\bibnamefont {Weinfurter}},\ and\ \bibinfo {author} {\bibfnamefont {A.}~\bibnamefont {Zeilinger}},\ }\bibfield  {title} {\bibinfo {title} {{A fast and compact quantum random number generator}},\ }\href {https://doi.org/10.1063/1.1150518} {\bibfield  {journal} {\bibinfo  {journal} {Review of Scientific Instruments}\ }\textbf {\bibinfo {volume} {71}},\ \bibinfo {pages} {1675} (\bibinfo {year} {2000})},\ \Eprint {https://arxiv.org/abs/https://pubs.aip.org/aip/rsi/article-pdf/71/4/1675/19183814/1675\_1\_online.pdf} {https://pubs.aip.org/aip/rsi/article-pdf/71/4/1675/19183814/1675\_1\_online.pdf} \BibitemShut {NoStop}%
\bibitem [{\citenamefont {Dynes}\ \emph {et~al.}(2008)\citenamefont {Dynes}, \citenamefont {Yuan}, \citenamefont {Sharpe},\ and\ \citenamefont {Shields}}]{SINGLE2}%
  \BibitemOpen
  \bibfield  {author} {\bibinfo {author} {\bibfnamefont {J.~F.}\ \bibnamefont {Dynes}}, \bibinfo {author} {\bibfnamefont {Z.~L.}\ \bibnamefont {Yuan}}, \bibinfo {author} {\bibfnamefont {A.~W.}\ \bibnamefont {Sharpe}},\ and\ \bibinfo {author} {\bibfnamefont {A.~J.}\ \bibnamefont {Shields}},\ }\bibfield  {title} {\bibinfo {title} {{A high speed, postprocessing free, quantum random number generator}},\ }\href {https://doi.org/10.1063/1.2961000} {\bibfield  {journal} {\bibinfo  {journal} {Applied Physics Letters}\ }\textbf {\bibinfo {volume} {93}},\ \bibinfo {pages} {031109} (\bibinfo {year} {2008})},\ \Eprint {https://arxiv.org/abs/https://pubs.aip.org/aip/apl/article-pdf/doi/10.1063/1.2961000/14398063/031109\_1\_online.pdf} {https://pubs.aip.org/aip/apl/article-pdf/doi/10.1063/1.2961000/14398063/031109\_1\_online.pdf} \BibitemShut {NoStop}%
\bibitem [{\citenamefont {Wayne}\ and\ \citenamefont {Kwiat}(2010)}]{SINGLE3}%
  \BibitemOpen
  \bibfield  {author} {\bibinfo {author} {\bibfnamefont {M.~A.}\ \bibnamefont {Wayne}}\ and\ \bibinfo {author} {\bibfnamefont {P.~G.}\ \bibnamefont {Kwiat}},\ }\bibfield  {title} {\bibinfo {title} {Low-bias high-speed quantum random number generator via shaped optical pulses},\ }\href {https://doi.org/10.1364/OE.18.009351} {\bibfield  {journal} {\bibinfo  {journal} {Opt. Express}\ }\textbf {\bibinfo {volume} {18}},\ \bibinfo {pages} {9351} (\bibinfo {year} {2010})}\BibitemShut {NoStop}%
\bibitem [{\citenamefont {F\"{u}rst}\ \emph {et~al.}(2010)\citenamefont {F\"{u}rst}, \citenamefont {Weier}, \citenamefont {Nauerth}, \citenamefont {Marangon}, \citenamefont {Kurtsiefer},\ and\ \citenamefont {Weinfurter}}]{SINGLE4}%
  \BibitemOpen
  \bibfield  {author} {\bibinfo {author} {\bibfnamefont {H.}~\bibnamefont {F\"{u}rst}}, \bibinfo {author} {\bibfnamefont {H.}~\bibnamefont {Weier}}, \bibinfo {author} {\bibfnamefont {S.}~\bibnamefont {Nauerth}}, \bibinfo {author} {\bibfnamefont {D.~G.}\ \bibnamefont {Marangon}}, \bibinfo {author} {\bibfnamefont {C.}~\bibnamefont {Kurtsiefer}},\ and\ \bibinfo {author} {\bibfnamefont {H.}~\bibnamefont {Weinfurter}},\ }\bibfield  {title} {\bibinfo {title} {High speed optical quantum random number generation},\ }\href {https://doi.org/10.1364/OE.18.013029} {\bibfield  {journal} {\bibinfo  {journal} {Opt. Express}\ }\textbf {\bibinfo {volume} {18}},\ \bibinfo {pages} {13029} (\bibinfo {year} {2010})}\BibitemShut {NoStop}%
\bibitem [{\citenamefont {Wahl}\ \emph {et~al.}(2011)\citenamefont {Wahl}, \citenamefont {Leifgen}, \citenamefont {Berlin}, \citenamefont {Röhlicke}, \citenamefont {Rahn},\ and\ \citenamefont {Benson}}]{SINGLE5}%
  \BibitemOpen
  \bibfield  {author} {\bibinfo {author} {\bibfnamefont {M.}~\bibnamefont {Wahl}}, \bibinfo {author} {\bibfnamefont {M.}~\bibnamefont {Leifgen}}, \bibinfo {author} {\bibfnamefont {M.}~\bibnamefont {Berlin}}, \bibinfo {author} {\bibfnamefont {T.}~\bibnamefont {Röhlicke}}, \bibinfo {author} {\bibfnamefont {H.-J.}\ \bibnamefont {Rahn}},\ and\ \bibinfo {author} {\bibfnamefont {O.}~\bibnamefont {Benson}},\ }\bibfield  {title} {\bibinfo {title} {{An ultrafast quantum random number generator with provably bounded output bias based on photon arrival time measurements}},\ }\href {https://doi.org/10.1063/1.3578456} {\bibfield  {journal} {\bibinfo  {journal} {Applied Physics Letters}\ }\textbf {\bibinfo {volume} {98}},\ \bibinfo {pages} {171105} (\bibinfo {year} {2011})},\ \Eprint {https://arxiv.org/abs/https://pubs.aip.org/aip/apl/article-pdf/doi/10.1063/1.3578456/14448338/171105\_1\_online.pdf} {https://pubs.aip.org/aip/apl/article-pdf/doi/10.1063/1.3578456/14448338/171105\_1\_online.pdf} \BibitemShut {NoStop}%
\bibitem [{\citenamefont {Collett}\ \emph {et~al.}(1987)\citenamefont {Collett}, \citenamefont {Loudon},\ and\ \citenamefont {Gardiner}}]{collett}%
  \BibitemOpen
  \bibfield  {author} {\bibinfo {author} {\bibfnamefont {M.}~\bibnamefont {Collett}}, \bibinfo {author} {\bibfnamefont {R.}~\bibnamefont {Loudon}},\ and\ \bibinfo {author} {\bibfnamefont {C.}~\bibnamefont {Gardiner}},\ }\bibfield  {title} {\bibinfo {title} {Quantum theory of optical homodyne and heterodyne detection},\ }\href {https://doi.org/10.1080/09500348714550811} {\bibfield  {journal} {\bibinfo  {journal} {Journal of Modern Optics}\ }\textbf {\bibinfo {volume} {34}},\ \bibinfo {pages} {881} (\bibinfo {year} {1987})},\ \Eprint {https://arxiv.org/abs/https://doi.org/10.1080/09500348714550811} {https://doi.org/10.1080/09500348714550811} \BibitemShut {NoStop}%
\bibitem [{\citenamefont {Yuen}\ and\ \citenamefont {Chan}(1983)}]{Yuen:83}%
  \BibitemOpen
  \bibfield  {author} {\bibinfo {author} {\bibfnamefont {H.~P.}\ \bibnamefont {Yuen}}\ and\ \bibinfo {author} {\bibfnamefont {V.~W.~S.}\ \bibnamefont {Chan}},\ }\bibfield  {title} {\bibinfo {title} {Noise in homodyne and heterodyne detection},\ }\href {https://doi.org/10.1364/OL.8.000177} {\bibfield  {journal} {\bibinfo  {journal} {Opt. Lett.}\ }\textbf {\bibinfo {volume} {8}},\ \bibinfo {pages} {177} (\bibinfo {year} {1983})}\BibitemShut {NoStop}%
\bibitem [{\citenamefont {Trifonov}\ \emph {et~al.}(2007)\citenamefont {Trifonov}, \citenamefont {Vig},\ and\ \citenamefont {Inc.}}]{Trifonov}%
  \BibitemOpen
  \bibfield  {author} {\bibinfo {author} {\bibfnamefont {A.}~\bibnamefont {Trifonov}}, \bibinfo {author} {\bibfnamefont {H.}~\bibnamefont {Vig}},\ and\ \bibinfo {author} {\bibfnamefont {M.~T.}\ \bibnamefont {Inc.}},\ }\bibfield  {title} {\bibinfo {title} {“quantum noise random number generator”},\ }\href@noop {} {\bibfield  {journal} {\bibinfo  {journal} {Patent US7284024}\ } (\bibinfo {year} {2007})}\BibitemShut {NoStop}%
\bibitem [{\citenamefont {Gabriel}\ \emph {et~al.}(2010)\citenamefont {Gabriel}, \citenamefont {Wittmann}, \citenamefont {Sych}, \citenamefont {Dong}, \citenamefont {Mauerer}, \citenamefont {Andersen}, \citenamefont {Marquardt},\ and\ \citenamefont {Leuchs}}]{Gabriel2010}%
  \BibitemOpen
  \bibfield  {author} {\bibinfo {author} {\bibfnamefont {C.}~\bibnamefont {Gabriel}}, \bibinfo {author} {\bibfnamefont {C.}~\bibnamefont {Wittmann}}, \bibinfo {author} {\bibfnamefont {D.}~\bibnamefont {Sych}}, \bibinfo {author} {\bibfnamefont {R.}~\bibnamefont {Dong}}, \bibinfo {author} {\bibfnamefont {W.}~\bibnamefont {Mauerer}}, \bibinfo {author} {\bibfnamefont {U.~L.}\ \bibnamefont {Andersen}}, \bibinfo {author} {\bibfnamefont {C.}~\bibnamefont {Marquardt}},\ and\ \bibinfo {author} {\bibfnamefont {G.}~\bibnamefont {Leuchs}},\ }\bibfield  {title} {\bibinfo {title} {A generator for unique quantum random numbers based on vacuum states},\ }\href {https://doi.org/10.1038/nphoton.2010.197} {\bibfield  {journal} {\bibinfo  {journal} {Nature Photonics}\ }\textbf {\bibinfo {volume} {4}},\ \bibinfo {pages} {711} (\bibinfo {year} {2010})}\BibitemShut {NoStop}%
\bibitem [{\citenamefont {Symul}\ \emph {et~al.}(2011)\citenamefont {Symul}, \citenamefont {Assad},\ and\ \citenamefont {Lam}}]{10.1063/1.3597793}%
  \BibitemOpen
  \bibfield  {author} {\bibinfo {author} {\bibfnamefont {T.}~\bibnamefont {Symul}}, \bibinfo {author} {\bibfnamefont {S.~M.}\ \bibnamefont {Assad}},\ and\ \bibinfo {author} {\bibfnamefont {P.~K.}\ \bibnamefont {Lam}},\ }\bibfield  {title} {\bibinfo {title} {{Real time demonstration of high bitrate quantum random number generation with coherent laser light}},\ }\href {https://doi.org/10.1063/1.3597793} {\bibfield  {journal} {\bibinfo  {journal} {Applied Physics Letters}\ }\textbf {\bibinfo {volume} {98}},\ \bibinfo {pages} {231103} (\bibinfo {year} {2011})},\ \Eprint {https://arxiv.org/abs/https://pubs.aip.org/aip/apl/article-pdf/doi/10.1063/1.3597793/14450740/231103\_1\_online.pdf} {https://pubs.aip.org/aip/apl/article-pdf/doi/10.1063/1.3597793/14450740/231103\_1\_online.pdf} \BibitemShut {NoStop}%
\bibitem [{\citenamefont {Haw}\ \emph {et~al.}(2015)\citenamefont {Haw}, \citenamefont {Assad}, \citenamefont {Lance}, \citenamefont {Ng}, \citenamefont {Sharma}, \citenamefont {Lam},\ and\ \citenamefont {Symul}}]{PhysRevApplied.3.054004}%
  \BibitemOpen
  \bibfield  {author} {\bibinfo {author} {\bibfnamefont {J.~Y.}\ \bibnamefont {Haw}}, \bibinfo {author} {\bibfnamefont {S.~M.}\ \bibnamefont {Assad}}, \bibinfo {author} {\bibfnamefont {A.~M.}\ \bibnamefont {Lance}}, \bibinfo {author} {\bibfnamefont {N.~H.~Y.}\ \bibnamefont {Ng}}, \bibinfo {author} {\bibfnamefont {V.}~\bibnamefont {Sharma}}, \bibinfo {author} {\bibfnamefont {P.~K.}\ \bibnamefont {Lam}},\ and\ \bibinfo {author} {\bibfnamefont {T.}~\bibnamefont {Symul}},\ }\bibfield  {title} {\bibinfo {title} {Maximization of extractable randomness in a quantum random-number generator},\ }\href {https://doi.org/10.1103/PhysRevApplied.3.054004} {\bibfield  {journal} {\bibinfo  {journal} {Phys. Rev. Appl.}\ }\textbf {\bibinfo {volume} {3}},\ \bibinfo {pages} {054004} (\bibinfo {year} {2015})}\BibitemShut {NoStop}%
\bibitem [{\citenamefont {Zhang}\ \emph {et~al.}(2016)\citenamefont {Zhang}, \citenamefont {Nie}, \citenamefont {Zhou}, \citenamefont {Liang}, \citenamefont {Ma}, \citenamefont {Zhang},\ and\ \citenamefont {Pan}}]{Zhang2016}%
  \BibitemOpen
  \bibfield  {author} {\bibinfo {author} {\bibfnamefont {X.-G.}\ \bibnamefont {Zhang}}, \bibinfo {author} {\bibfnamefont {Y.-Q.}\ \bibnamefont {Nie}}, \bibinfo {author} {\bibfnamefont {H.}~\bibnamefont {Zhou}}, \bibinfo {author} {\bibfnamefont {H.}~\bibnamefont {Liang}}, \bibinfo {author} {\bibfnamefont {X.}~\bibnamefont {Ma}}, \bibinfo {author} {\bibfnamefont {J.}~\bibnamefont {Zhang}},\ and\ \bibinfo {author} {\bibfnamefont {J.-W.}\ \bibnamefont {Pan}},\ }\bibfield  {title} {\bibinfo {title} {{Note: Fully integrated 3.2 Gbps quantum random number generator with real-time extraction}},\ }\href {https://doi.org/10.1063/1.4958663} {\bibfield  {journal} {\bibinfo  {journal} {Review of Scientific Instruments}\ }\textbf {\bibinfo {volume} {87}},\ \bibinfo {pages} {076102} (\bibinfo {year} {2016})},\ \Eprint {https://arxiv.org/abs/https://pubs.aip.org/aip/rsi/article-pdf/doi/10.1063/1.4958663/14737785/076102\_1\_online.pdf}
  {https://pubs.aip.org/aip/rsi/article-pdf/doi/10.1063/1.4958663/14737785/076102\_1\_online.pdf} \BibitemShut {NoStop}%
\bibitem [{\citenamefont {Zheng}\ \emph {et~al.}(2019)\citenamefont {Zheng}, \citenamefont {Zhang}, \citenamefont {Huang}, \citenamefont {Yu},\ and\ \citenamefont {Guo}}]{10.1063/1.5078547}%
  \BibitemOpen
  \bibfield  {author} {\bibinfo {author} {\bibfnamefont {Z.}~\bibnamefont {Zheng}}, \bibinfo {author} {\bibfnamefont {Y.}~\bibnamefont {Zhang}}, \bibinfo {author} {\bibfnamefont {W.}~\bibnamefont {Huang}}, \bibinfo {author} {\bibfnamefont {S.}~\bibnamefont {Yu}},\ and\ \bibinfo {author} {\bibfnamefont {H.}~\bibnamefont {Guo}},\ }\bibfield  {title} {\bibinfo {title} {{6 Gbps real-time optical quantum random number generator based on vacuum fluctuation}},\ }\href {https://doi.org/10.1063/1.5078547} {\bibfield  {journal} {\bibinfo  {journal} {Review of Scientific Instruments}\ }\textbf {\bibinfo {volume} {90}},\ \bibinfo {pages} {043105} (\bibinfo {year} {2019})},\ \Eprint {https://arxiv.org/abs/https://pubs.aip.org/aip/rsi/article-pdf/doi/10.1063/1.5078547/16012915/043105\_1\_online.pdf} {https://pubs.aip.org/aip/rsi/article-pdf/doi/10.1063/1.5078547/16012915/043105\_1\_online.pdf} \BibitemShut {NoStop}%
\bibitem [{\citenamefont {Marangon}\ \emph {et~al.}(2017)\citenamefont {Marangon}, \citenamefont {Vallone},\ and\ \citenamefont {Villoresi}}]{PhysRevLett.118.060503}%
  \BibitemOpen
  \bibfield  {author} {\bibinfo {author} {\bibfnamefont {D.~G.}\ \bibnamefont {Marangon}}, \bibinfo {author} {\bibfnamefont {G.}~\bibnamefont {Vallone}},\ and\ \bibinfo {author} {\bibfnamefont {P.}~\bibnamefont {Villoresi}},\ }\bibfield  {title} {\bibinfo {title} {Source-device-independent ultrafast quantum random number generation},\ }\href {https://doi.org/10.1103/PhysRevLett.118.060503} {\bibfield  {journal} {\bibinfo  {journal} {Phys. Rev. Lett.}\ }\textbf {\bibinfo {volume} {118}},\ \bibinfo {pages} {060503} (\bibinfo {year} {2017})}\BibitemShut {NoStop}%
\bibitem [{\citenamefont {Michel}\ \emph {et~al.}(2019)\citenamefont {Michel}, \citenamefont {Haw}, \citenamefont {Marangon}, \citenamefont {Thearle}, \citenamefont {Vallone}, \citenamefont {Villoresi}, \citenamefont {Lam},\ and\ \citenamefont {Assad}}]{PhysRevApplied.12.034017}%
  \BibitemOpen
  \bibfield  {author} {\bibinfo {author} {\bibfnamefont {T.}~\bibnamefont {Michel}}, \bibinfo {author} {\bibfnamefont {J.~Y.}\ \bibnamefont {Haw}}, \bibinfo {author} {\bibfnamefont {D.~G.}\ \bibnamefont {Marangon}}, \bibinfo {author} {\bibfnamefont {O.}~\bibnamefont {Thearle}}, \bibinfo {author} {\bibfnamefont {G.}~\bibnamefont {Vallone}}, \bibinfo {author} {\bibfnamefont {P.}~\bibnamefont {Villoresi}}, \bibinfo {author} {\bibfnamefont {P.~K.}\ \bibnamefont {Lam}},\ and\ \bibinfo {author} {\bibfnamefont {S.~M.}\ \bibnamefont {Assad}},\ }\bibfield  {title} {\bibinfo {title} {Real-time source-independent quantum random-number generator with squeezed states},\ }\href {https://doi.org/10.1103/PhysRevApplied.12.034017} {\bibfield  {journal} {\bibinfo  {journal} {Phys. Rev. Appl.}\ }\textbf {\bibinfo {volume} {12}},\ \bibinfo {pages} {034017} (\bibinfo {year} {2019})}\BibitemShut {NoStop}%
\bibitem [{\citenamefont {Avesani}\ \emph {et~al.}(2018)\citenamefont {Avesani}, \citenamefont {Marangon}, \citenamefont {Vallone},\ and\ \citenamefont {Villoresi}}]{Avesani2018}%
  \BibitemOpen
  \bibfield  {author} {\bibinfo {author} {\bibfnamefont {M.}~\bibnamefont {Avesani}}, \bibinfo {author} {\bibfnamefont {D.~G.}\ \bibnamefont {Marangon}}, \bibinfo {author} {\bibfnamefont {G.}~\bibnamefont {Vallone}},\ and\ \bibinfo {author} {\bibfnamefont {P.}~\bibnamefont {Villoresi}},\ }\bibfield  {title} {\bibinfo {title} {Source-device-independent heterodyne-based quantum random number generator at 17 gbps},\ }\href {https://doi.org/10.1038/s41467-018-07585-0} {\bibfield  {journal} {\bibinfo  {journal} {Nature Communications}\ }\textbf {\bibinfo {volume} {9}},\ \bibinfo {pages} {5365} (\bibinfo {year} {2018})}\BibitemShut {NoStop}%
\bibitem [{\citenamefont {Xu}\ \emph {et~al.}(2019)\citenamefont {Xu}, \citenamefont {Chen}, \citenamefont {Li}, \citenamefont {Yang}, \citenamefont {Su}, \citenamefont {Huang}, \citenamefont {Zhang},\ and\ \citenamefont {Guo}}]{Xu_2019}%
  \BibitemOpen
  \bibfield  {author} {\bibinfo {author} {\bibfnamefont {B.}~\bibnamefont {Xu}}, \bibinfo {author} {\bibfnamefont {Z.}~\bibnamefont {Chen}}, \bibinfo {author} {\bibfnamefont {Z.}~\bibnamefont {Li}}, \bibinfo {author} {\bibfnamefont {J.}~\bibnamefont {Yang}}, \bibinfo {author} {\bibfnamefont {Q.}~\bibnamefont {Su}}, \bibinfo {author} {\bibfnamefont {W.}~\bibnamefont {Huang}}, \bibinfo {author} {\bibfnamefont {Y.}~\bibnamefont {Zhang}},\ and\ \bibinfo {author} {\bibfnamefont {H.}~\bibnamefont {Guo}},\ }\bibfield  {title} {\bibinfo {title} {High speed continuous variable source-independent quantum random number generation},\ }\href {https://doi.org/10.1088/2058-9565/ab0fd9} {\bibfield  {journal} {\bibinfo  {journal} {Quantum Science and Technology}\ }\textbf {\bibinfo {volume} {4}},\ \bibinfo {pages} {025013} (\bibinfo {year} {2019})}\BibitemShut {NoStop}%
\bibitem [{\citenamefont {Smith}\ \emph {et~al.}(2019)\citenamefont {Smith}, \citenamefont {Marangon}, \citenamefont {Lucamarini}, \citenamefont {Yuan},\ and\ \citenamefont {Shields}}]{PhysRevA.99.062326}%
  \BibitemOpen
  \bibfield  {author} {\bibinfo {author} {\bibfnamefont {P.~R.}\ \bibnamefont {Smith}}, \bibinfo {author} {\bibfnamefont {D.~G.}\ \bibnamefont {Marangon}}, \bibinfo {author} {\bibfnamefont {M.}~\bibnamefont {Lucamarini}}, \bibinfo {author} {\bibfnamefont {Z.~L.}\ \bibnamefont {Yuan}},\ and\ \bibinfo {author} {\bibfnamefont {A.~J.}\ \bibnamefont {Shields}},\ }\bibfield  {title} {\bibinfo {title} {Simple source device-independent continuous-variable quantum random number generator},\ }\href {https://doi.org/10.1103/PhysRevA.99.062326} {\bibfield  {journal} {\bibinfo  {journal} {Phys. Rev. A}\ }\textbf {\bibinfo {volume} {99}},\ \bibinfo {pages} {062326} (\bibinfo {year} {2019})}\BibitemShut {NoStop}%
\bibitem [{\citenamefont {Cheng}\ \emph {et~al.}(2022)\citenamefont {Cheng}, \citenamefont {Qin}, \citenamefont {Liang}, \citenamefont {Li}, \citenamefont {Yan}, \citenamefont {Jia},\ and\ \citenamefont {Peng}}]{Cheng:22}%
  \BibitemOpen
  \bibfield  {author} {\bibinfo {author} {\bibfnamefont {J.}~\bibnamefont {Cheng}}, \bibinfo {author} {\bibfnamefont {J.}~\bibnamefont {Qin}}, \bibinfo {author} {\bibfnamefont {S.}~\bibnamefont {Liang}}, \bibinfo {author} {\bibfnamefont {J.}~\bibnamefont {Li}}, \bibinfo {author} {\bibfnamefont {Z.}~\bibnamefont {Yan}}, \bibinfo {author} {\bibfnamefont {X.}~\bibnamefont {Jia}},\ and\ \bibinfo {author} {\bibfnamefont {K.}~\bibnamefont {Peng}},\ }\bibfield  {title} {\bibinfo {title} {Mutually testing source-device-independent quantum random number generator},\ }\href {https://doi.org/10.1364/PRJ.444853} {\bibfield  {journal} {\bibinfo  {journal} {Photon. Res.}\ }\textbf {\bibinfo {volume} {10}},\ \bibinfo {pages} {646} (\bibinfo {year} {2022})}\BibitemShut {NoStop}%
\bibitem [{\citenamefont {Zhang}\ \emph {et~al.}(2018)\citenamefont {Zhang}, \citenamefont {Zhang}, \citenamefont {Li}, \citenamefont {Yu},\ and\ \citenamefont {Guo}}]{8444971}%
  \BibitemOpen
  \bibfield  {author} {\bibinfo {author} {\bibfnamefont {X.}~\bibnamefont {Zhang}}, \bibinfo {author} {\bibfnamefont {Y.}~\bibnamefont {Zhang}}, \bibinfo {author} {\bibfnamefont {Z.}~\bibnamefont {Li}}, \bibinfo {author} {\bibfnamefont {S.}~\bibnamefont {Yu}},\ and\ \bibinfo {author} {\bibfnamefont {H.}~\bibnamefont {Guo}},\ }\bibfield  {title} {\bibinfo {title} {1.2-ghz balanced homodyne detector for continuous-variable quantum information technology},\ }\href {https://doi.org/10.1109/JPHOT.2018.2866514} {\bibfield  {journal} {\bibinfo  {journal} {IEEE Photonics Journal}\ }\textbf {\bibinfo {volume} {10}},\ \bibinfo {pages} {1} (\bibinfo {year} {2018})}\BibitemShut {NoStop}%
\bibitem [{\citenamefont {Wolf}\ \emph {et~al.}(2006)\citenamefont {Wolf}, \citenamefont {Giedke},\ and\ \citenamefont {Cirac}}]{PhysRevLett.96.080502}%
  \BibitemOpen
  \bibfield  {author} {\bibinfo {author} {\bibfnamefont {M.~M.}\ \bibnamefont {Wolf}}, \bibinfo {author} {\bibfnamefont {G.}~\bibnamefont {Giedke}},\ and\ \bibinfo {author} {\bibfnamefont {J.~I.}\ \bibnamefont {Cirac}},\ }\bibfield  {title} {\bibinfo {title} {Extremality of gaussian quantum states},\ }\href {https://doi.org/10.1103/PhysRevLett.96.080502} {\bibfield  {journal} {\bibinfo  {journal} {Phys. Rev. Lett.}\ }\textbf {\bibinfo {volume} {96}},\ \bibinfo {pages} {080502} (\bibinfo {year} {2006})}\BibitemShut {NoStop}%
\bibitem [{\citenamefont {Garc\'{\i}a-Patr\'on}\ and\ \citenamefont {Cerf}(2006)}]{PhysRevLett.97.190503}%
  \BibitemOpen
  \bibfield  {author} {\bibinfo {author} {\bibfnamefont {R.}~\bibnamefont {Garc\'{\i}a-Patr\'on}}\ and\ \bibinfo {author} {\bibfnamefont {N.~J.}\ \bibnamefont {Cerf}},\ }\bibfield  {title} {\bibinfo {title} {Unconditional optimality of gaussian attacks against continuous-variable quantum key distribution},\ }\href {https://doi.org/10.1103/PhysRevLett.97.190503} {\bibfield  {journal} {\bibinfo  {journal} {Phys. Rev. Lett.}\ }\textbf {\bibinfo {volume} {97}},\ \bibinfo {pages} {190503} (\bibinfo {year} {2006})}\BibitemShut {NoStop}%
\bibitem [{\citenamefont {Bassham}\ \emph {et~al.}(2010)\citenamefont {Bassham}, \citenamefont {Rukhin}, \citenamefont {Soto}, \citenamefont {Nechvatal}, \citenamefont {Smid}, \citenamefont {Leigh}, \citenamefont {Levenson}, \citenamefont {Vangel}, \citenamefont {Heckert},\ and\ \citenamefont {Banks}}]{lawrence}%
  \BibitemOpen
  \bibfield  {author} {\bibinfo {author} {\bibfnamefont {L.}~\bibnamefont {Bassham}}, \bibinfo {author} {\bibfnamefont {A.}~\bibnamefont {Rukhin}}, \bibinfo {author} {\bibfnamefont {J.}~\bibnamefont {Soto}}, \bibinfo {author} {\bibfnamefont {J.}~\bibnamefont {Nechvatal}}, \bibinfo {author} {\bibfnamefont {M.}~\bibnamefont {Smid}}, \bibinfo {author} {\bibfnamefont {S.}~\bibnamefont {Leigh}}, \bibinfo {author} {\bibfnamefont {M.}~\bibnamefont {Levenson}}, \bibinfo {author} {\bibfnamefont {M.}~\bibnamefont {Vangel}}, \bibinfo {author} {\bibfnamefont {N.}~\bibnamefont {Heckert}},\ and\ \bibinfo {author} {\bibfnamefont {D.}~\bibnamefont {Banks}},\ }\bibfield  {title} {\bibinfo {title} {A statistical test suite for random and pseudorandom number generators for cryptographic applications},\ }\href@noop {} {\  (\bibinfo {year} {2010})}\BibitemShut {NoStop}%
\end{thebibliography}%

\end{document}